\documentclass[final,5p,times,twocolumn]{elsarticle}

\usepackage{amssymb}
\usepackage{lipsum}
\usepackage{adjustbox} 
\usepackage{caption} 
\usepackage{multirow}
\usepackage{booktabs}
\usepackage{float}
\usepackage{placeins}
\setcitestyle{square}
\usepackage[numbers]{natbib}
\usepackage{siunitx}
\usepackage{xcolor}
\usepackage{hyperref}
\usepackage{subcaption}
\usepackage{amsmath}
\usepackage{cleveref}
\usepackage{soul}
\usepackage{graphicx}
\usepackage{flushend}

\newcommand{\hertz}{\SI{}{\hertz}}

\journal{Astronomy and Computing}

\begin{document}

\begin{frontmatter}

\title{Binary Black Hole Parameter Estimation with Hybrid CNN-Transformer Neural Networks}

\author[first]{Panagiotis N. Sakellariou}
\ead{psakellariou@uth.gr}
\author[first]{Spiros V. Georgakopoulos}
\ead{spirosgeorg@uth.gr}
\affiliation[first]{organization={Department of Mathematics, University of Thessaly},
            addressline={3rd km. Old National Road Lamia-Athens}, 
            city={Lamia},
            postcode={35100}, 
            country={Greece}}
\author[2]{Sotiris Tasoulis}
\ead{stasoulis@uth.gr}

\affiliation[2]{organization={Department of Computer Science and Biomedical Informatics, University of Thessaly}, 
                postcode={35100}, 
                country={Greece}}

\author[2]{Vassilis P. Plagianakos}
\ead{vpp@uth.gr}

\begin{abstract}
    The detection of gravitational waves has revolutionized our ability to explore fundamental aspects of the Universe. Traditionally, modeled gravitational-wave signals have been identified using template-based matched filtering, followed by coincidence analysis across multiple detectors in the signal-to-noise ratio time series. Recent advances in Machine Learning and Deep Learning have sparked growing interest in their application to both signal detection and parameter estimation. 
    In this study, a hybrid Deep Learning strategy is proposed that leverages the effectiveness of Transformer encoders alongside well-established Convolutional Neural Network architectures in an attempt to estimate the intrinsic and extrinsic parameters of non-precessing binary black hole systems. The primary focus of this work is point estimation, producing single best-fit values for each parameter rather than full posterior distributions. This method is evaluated on both simulated signals embedded in Gaussian noise and real gravitational-wave events, and it demonstrates strong predictive performance and robustness across key astrophysical parameters.

\end{abstract}


\end{frontmatter}

\section{Introduction}
\label{introduction}
Gravitational waves are ripples in the fabric of space-time, produced by some of the most extreme and energetic events in the universe. Their existence was first predicted by Albert Einstein in 1916 as a consequence of his general theory of relativity. According to his equations, accelerating massive objects, such as merging black holes or neutron stars, generate disturbances that propagate outward as waves. These waves travel at the speed of light, carrying crucial information about their sources and offering insights into the fundamental nature of gravity.

The first direct detection of gravitational waves occurred on September 14, 2015, when the Laser Interferometer Gravitational-Wave Observatory (LIGO) \cite{aasi_advanced_2015} observed a signal from a Binary Black Hole (BBH) merger \cite{abbott_observation_2016}. Since then, LIGO and Virgo \cite{acernese_advanced_2014} have consistently detected signals from Compact Binary Coalescences (CBCs) over three observational runs \cite{abbott_gwtc-1_2019, abbott_gwtc-2_2021, collaboration_gwtc-21_2022, abbott_gwtc-3_2023}. A fourth run is now underway, incorporating the Kamioka Gravitational Wave Detector (KAGRA) \cite{akutsu_overview_2020}. As these detectors undergo further upgrades to enhance their sensitivity, and with new facilities such as LIGO-India \cite{saleem_science_2021} set to join the network, the rate of gravitational wave detections is expected to increase significantly.

The LIGO-Virgo Collaboration (LVC) has identified numerous gravitational wave events, most of which originate from BBH mergers. BBHs emit gravitational waves during their in-spiral, merger, and ring-down phases. However, mergers involving Binary Neutron Stars (BNS) and Neutron Star–Black Hole (NSBH) systems have also been observed. Rapid detection of these events is essential, as it enables immediate follow-up observations with electromagnetic telescopes. To facilitate this, fast parameter estimation is required to assess the probability that a merger involves a neutron star, which is more likely to produce an accompanying electromagnetic counterpart.

Compact binary mergers are fully described by a 15-dimensional vector, encompassing both intrinsic properties (such as mass and spin) and extrinsic factors (such as the system’s orientation relative to the detector). When a gravitational wave signal is detected, parameter estimation is performed to extract these properties.

Currently, CBC search pipelines rely on matched filtering to identify gravitational wave events in real time, while parameter estimation is conducted using Bayesian inference techniques such as Markov Chain Monte Carlo (MCMC) and nested sampling. Although these methods yield highly accurate results, they are computationally expensive, often requiring hours or even days to fully analyze a BBH merger. This delay poses a challenge for BNS mergers, where electromagnetic radiation may be emitted just seconds after the event. Consequently, rapid parameter estimation is crucial for capturing these transient electromagnetic signals in real time.

Deep Learning, a specialized branch of Machine Learning (ML), utilizes Artificial Neural Networks to analyze data and extract meaningful features for tasks such as classification and regression. In recent years, it has become a powerful tool in gravitational-wave astronomy, enhancing detection sensitivity, accelerating parameter estimation, and aiding in source classification and sky localization. Early Deep Learning applications in gravitational wave astronomy introduced Convolutional Neural Network (CNNs) based models for real-time detection and mass estimation of BBH systems \cite{george_deep_2018, george_deep_2018-1, gabbard_matching_2018, wei_deep_2021}, achieving performance comparable to matched filtering on real LIGO data. Subsequent studies expanded these methods to improve signal–noise discrimination and parameter estimation in more realistic settings \cite{krastev_real-time_2020, krastev_detection_2021, shen_statistically-informed_2021, narola_relative_2023, tang_deep_2024}. Likelihood-free inference approaches using neural posterior estimation and normalizing flows have further enabled rapid, high-dimensional parameter estimation for BBH and BNS systems \cite{Green_2020, green2020completeparameterinferencegw150914, Dax_2021, Dax_2025}. Deep Learning has also been applied to source classification, sky localization, and signal reconstruction, facilitating real-time alerts and interpretation of complex signals such as microlensed or premerger events \cite{chatterjee_using_2019, chatterjee_rapid_2023, chatterjee_premerger_2023, qiu_deep_2023, kim_deep_2022, nerin_parameter_2024, raza_explaining_2023, raza_gwskynet-multi_2025, dooney_deepextractor_2025}. Compared to traditional Bayesian inference methods, Deep Learning approaches offer the promise of significantly reducing computational time, making them particularly valuable for time-sensitive gravitational-wave analyses.

In this work, a hybrid Deep Learning approach is proposed for rapid point estimation (regression) of both intrinsic and extrinsic parameters of BBH systems from their gravitational-wave signals, enabling significantly faster inference compared to traditional Bayesian methods. BBH systems are used as a foundational test case to validate the approach, which can later be extended to BNS and NSBH systems, where rapid and accurate parameter estimation is critical for multi-messenger follow-up observations. Unlike probabilistic inference methods, which aim to recover full posterior distributions, the primary focus of this work is on producing single best-fit estimates. However, to further illustrate the advantage of the hybrid approach, it is also combined with normalizing flows to obtain posterior estimates, demonstrating the superiority of the learned feature representation compared to CNN-only models. The approach integrates CNNs and Transformer Encoder layers to effectively capture both local waveform patterns and long-range dependencies within noisy time-series data.

Convolutional layers are employed to impose an inductive bias of locality and shift-invariance, making them effective for extracting and denoising local features in the signal \cite{GEORGAKOPOULOS201823}. However, CNNs are inherently limited in capturing long-range temporal dependencies \cite{Jiang_2023}, which are essential for modeling the complete evolution of a BBH coalescence, from the inspiral to merger and ringdown phases. To overcome this limitation, self-attention mechanisms from the Transformer architecture are incorporated to dynamically assign importance to different time steps based on their relevance to the prediction task. This synergy enables more accurate parameter estimation than either component could achieve independently~\cite{ElZaar2025Hybrid}. Local features are first extracted and denoised through convolutional operations, forming a compact “vocabulary” of signal components. These are then interpreted by the self-attention layers to identify globally important patterns across the signal.

The methodology involves applying this hybrid architecture to estimate six physical parameters: the component masses and z-axis spin components of the two black holes, the luminosity distance of the merger event, and the inclination angle relative to the detector. To simulate realistic detector conditions, the input signals are embedded in Gaussian noise representative of the LIGO detectors’ sensitivity.

The remainder of this work is structured as follows: Section 2 presents the methodology, including data generation and the hybrid modeling approach. Section 3 reports the experimental results and discusses the findings. Section 4 summarizes our conclusions and outlines the broader impact of this work.

\section{Methodology}
\label{methodology}
This section introduces the hybrid CNN–Transformer approach for parameter estimation in non-precessing BBH systems using gravitational wave time-series data.

\subsection{Data Generation}
To achieve our research objectives, an initial approach was to use recorded gravitational wave strains from real events cataloged in the Gravitational Wave Open Science Center (GWOSC) as training data for our models. However, it was found that the available dataset was insufficient for adequately training the complex ML systems. Consequently, synthetic time-domain waveforms representing gravitational waves were generated using Effective-One-Body (EOB) Numerical-Relativity (NR) waveform models, specifically designed for spin-aligned binary black hole systems, known as SEOBNR models.

In this study, the SEOBNRv4 \cite{bohe_improved_2017} algorithm is employed to generate both the plus and cross polarization components of gravitational waves. These dimensionless polarizations are analogous to those of electromagnetic waves, but differ by 45 degrees rather than 90. The resulting waveforms are then projected onto the LIGO Hanford detector (H1) response using the PyCBC library \cite{nitz_gwastropycbc_2023}, yielding a single strain that represents the detector output.

The dataset used for training consists of time-domain gravitational wave strains sampled at $8192 \hertz{}$, with a low-frequency cutoff of $10 \hertz{}$ $(f_{low}=10 \hertz{})$. Each input to the network corresponds to a single projected strain signal, represented as a one-dimensional array of length 8192, where the single channel corresponds to the detector strain. These waveforms originate from non-precessing BBH systems with individual component masses ranging in integer steps of one within the range of $5M_\odot$–$100M_\odot$.

To prevent duplicate events, a constraint is imposed such that the primary mass ($m_1$) is always equal to or greater than the secondary mass ($m_2$). Additionally, the luminosity distance ($d_{L}$) of the generated gravitational wave signals is systematically varied from 50Mpc to 500Mpc in increments of 50Mpc to facilitate accurate distance estimation. The inclination angle ($\iota$) of the system is restricted to discrete values of $0$, $\pi/6$, $\pi/4$, $\pi/3$, and $\pi/2$ radians, which are converted to degrees before being fed into the Neural Networks. Finally, the dimensionless spin components along the z-axis of the two black holes ($\chi_{z_{1}}$, $\chi_{z_{2}}$) are varied from -1 to 1 in steps of 0.5, where -1 corresponds to the maximum spin anti-aligned with the orbital angular momentum.

Since different mass combinations produce gravitational waves of varying durations, a temporal window capturing the final second of each strain is applied. Consequently, all strains are standardized to 8192 values. If the strain duration is shorter than one second, zero-padding is applied at the end to ensure a uniform length.

In this work, the strain amplitude of the gravitational waves lies between $10^{-23}-10^{-20}$. These extremely small values can lead to vanishing gradients during the application of the backpropagation algorithm, making Neural Network training inefficient. To mitigate this, min-max normalization is applied, scaling the values to the range $[0,1]$. This transformation helps maintain numerical stability and improves the learning process of the Neural Network. 

The final dataset \footnote{The dataset and code will be released upon publication acceptance. \href{https://github.com/PanagiotisSakellariou/hybrid-bbh-parameter-estimation}{GitHub repository}.} consists of 5,820,000 samples (355.783 GB), randomly partitioned as follows: $30\%$ (1,746,000 samples, 106.735 GB) for testing, and the remaining $70\%$ split into $90\%$ for training (3,666,600 samples, 224.143 GB) and $10\%$ for validation (407,000 samples, 24.905 GB).

\begin{figure}[H]
  \centering
  \includegraphics[width=0.85\linewidth]{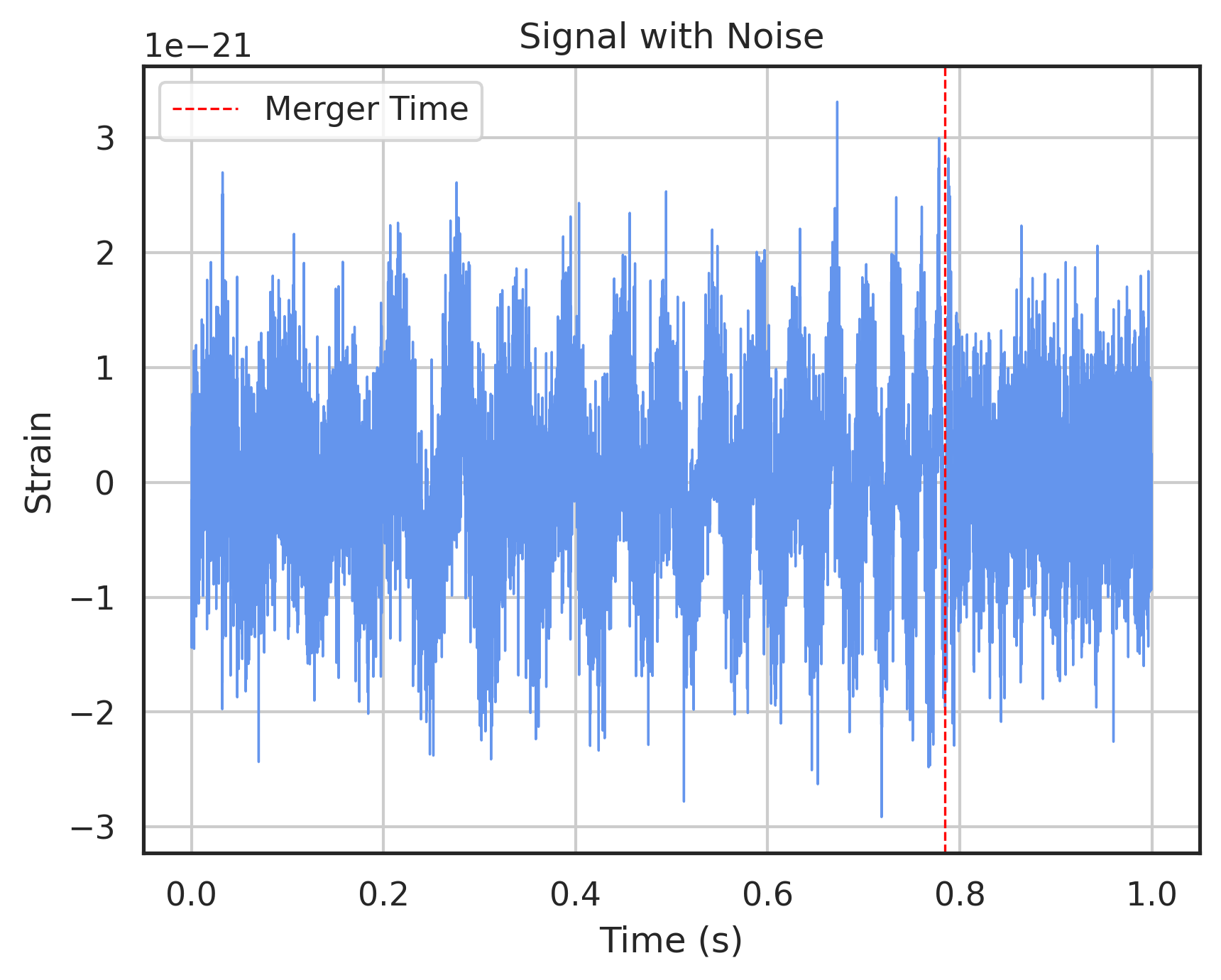}
  \caption{The resulting signal from a gravitational wave emitted by a BBH system with component masses of $80M_\odot$ and $30M_\odot$, located at a luminosity distance of 500Mpc, that was injected into Gaussian noise resembling the LIGO Hanford (H1) detector environment. The red line indicates the time of the merger phase.}
  \label{fig:signalwithnoise}
\end{figure}

To better approximate real observational data, the gravitational wave signals are embedded in Gaussian noise that simulates the noise characteristics of the LIGO detector.  ~\autoref{fig:signalwithnoise} presents an example of a strain as observed by the H1 detector after noise has been added, illustrating the numerical input data used for the neural networks. The only difference is that the signal in the figure is not min-max normalized. This example directly reflects the data generation procedure employed in this work.

\subsection{Hybrid CNN-Transformer Approach}
The general approach explored in this work involves enhancing parameter estimation from gravitational wave time-series data by integrating CNNs with Transformer Encoder layers. CNNs are used to extract localized features from raw strain data, while the self-attention mechanism in Transformers captures long-range correlations within the signal. This combination is designed to exploit the strengths of both architectures for time-series modeling.

This hybrid strategy is applied to the task of estimating six physical parameters from gravitational-wave signals. It was observed that the inclusion of Transformer layers consistently improves the performance of the CNN.

The proposed pipeline consists of a series of convolutional and pooling layers followed by Transformer Encoder layers and dense layers. Let $x \in \mathbb{R}^N$ denote the raw gravitational wave strain sampled over $N$ time steps. The convolutional component extracts local features, producing:

\begin{equation}
h^{(L)} =f_{\mathrm{conv}}(x)\in \mathbb{R}^{N'\times d},
\qquad
N' =\frac{N}{\prod_{\ell=1}^L p_\ell},
\end{equation}

\noindent where $p_\ell$ is the pooling factor at layer $\ell$, and $d$ is the hidden feature dimension. Each convolutional layer is followed by a non-linear activation and a pooling operation:

\begin{equation}
h^{(\ell)}
=\mathrm{Pool}_{p_\ell}\!\Bigl(\,
\sigma\bigl(\mathrm{Conv1d}(h^{(\ell-1)} W^{(\ell)},b^{(\ell)})\bigr)
\Bigr),
\quad
h^{(0)}=x,
\end{equation}

\noindent Following convolution and pooling, global dependencies are modeled through a self-attention mechanism applied to the resulting features:

\begin{equation}
\mathrm{Att}(X)
=\mathrm{softmax} \Bigl(QK^\top/\sqrt{d_k}\Bigr)\,V,
\quad
\end{equation}
\begin{equation}
Q = XW_Q,\;K = XW_K,\;V = XW_V,
\quad
X\in\mathbb{R}^{N\times d},
\end{equation}

\noindent By downsampling the sequence length from $N$ to $N'$, the quadratic cost of attention is significantly reduced, achieving computational efficiency without sacrificing expressivity. The overall abstract structure of the proposed method is expressed as:

\begin{equation}
\underbrace{x\in \mathbb{R}^N}_{\text{raw strain}} \xrightarrow{\text{Conv + Pool}} \underbrace{h\in \mathbb{R}^{N' \times d}} \xrightarrow{\text{Transformer}} \underbrace{z\in \mathbb{R}^{N' \times d}}_{\text{global context}} \xrightarrow{\text{Linear}} \underbrace{\hat{y}\in \mathbb{R}^6}
\end{equation}

\begin{figure*}[]
   \centering
   \includegraphics[width=1\linewidth]{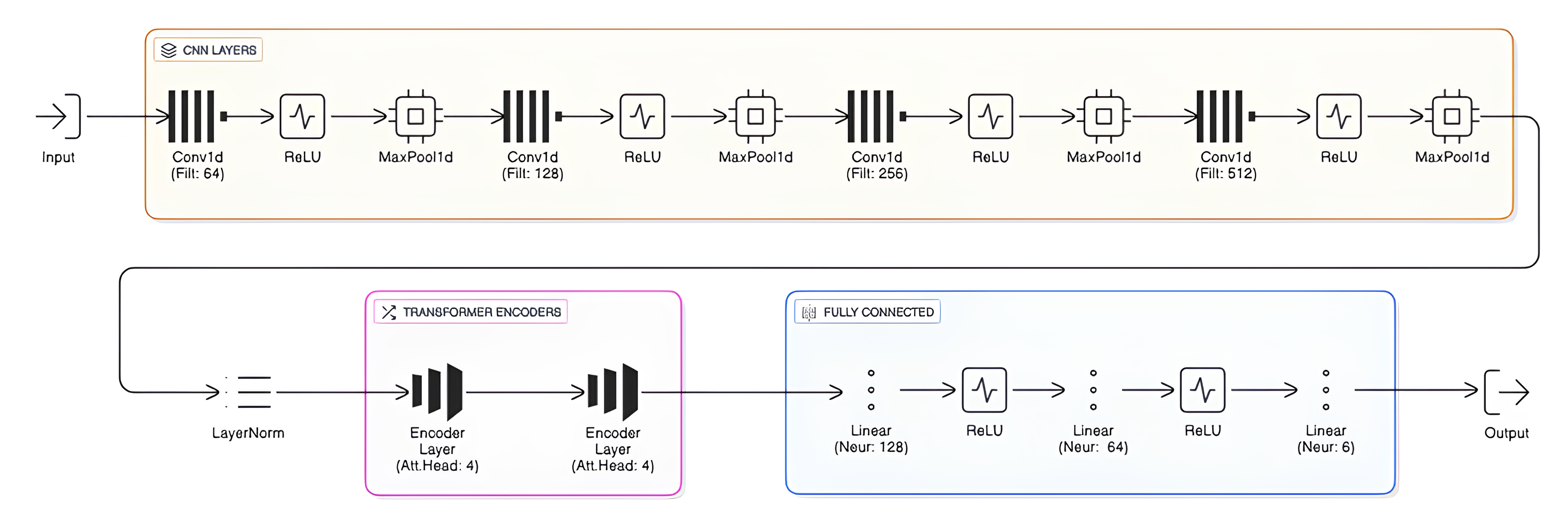}
   \caption{Schematic representation of the DeepModel-Hybrid architecture. The model combines CNN layers for local feature extraction with Transformer encoder layers for capturing long-range dependencies in the gravitational wave strain data. The final fully connected layers map the learned features to the target physical parameters.}
   \label{fig:hybrid-architecture}
\end{figure*}

\begin{table}[h!]
\centering
\caption{The architecture and the number of parameters of the DeepModel-Hybrid.}
\resizebox{0.85\linewidth}{!}{%
    \begin{tabular}{lrr}
        \multicolumn{3}{c}{\textbf{DeepModel-Hybrid}}                      \\ \hline
        {\textbf{Layer (type)}} & {\textbf{Output Shape}} & {\textbf{Param}} \\ \hline
        {\textbf{Input (tensor)}}& {(1, 8192)}            & {0}               \\
        {\textbf{Conv1d}}       & {(64, 8177)}            & {1,088}           \\
        {\textbf{MaxPooling1d}} & {(64, 2044)}            & {0}              \\
        {\textbf{Conv1d}}       & {(128, 2014)}           & {131,200}         \\
        {\textbf{MaxPooling1d}} & {(128, 503)}            & {0}              \\
        {\textbf{Conv1d}}       & {(256, 473)}            & {524,544}           \\
        {\textbf{MaxPooling1d}} & {(256, 118)}            & {0}              \\
        {\textbf{Conv1d}}       & {(512, 56)}            & {4,194,816}           \\
        {\textbf{MaxPooling1d}} & {(512, 14)}            & {0}              \\
        {\textbf{LayerNorm}}    & {(14, 512)}            & {1024}            \\
        {\textbf{TransformerEncoder}}       & {(14, 512)}            & {3,152,384}         \\
        {\textbf{TransformerEncoder}} & {(14, 512)}            & {3,152,384}              \\
        {\textbf{Dense}}        & {(128)}                  & {65,664}           \\
        {\textbf{Dense}}        & {(64)}                  & {8,256}           \\
        {\textbf{Dense}}        & {(6)}                   & {390}            \\ \hline
        \multicolumn{3}{l}{\textbf{Total params:} 11,231,750}                                   \\ \hline
    \end{tabular}
}
\label{table:proposed}
\end{table}

To demonstrate the proposed approach in practice, a specific implementation referred to as DeepModel-Hybrid is utilized. This model implements the proposed hybrid scheme using a specific stack of convolutional and pooling layers~\cite{george_deep_2018-1}, followed by two Transformer Encoder layers and three fully connected layers.

The CNN feature extractor consists of four one-dimensional convolutional layers interleaved with max-pooling layers. The first one-dimensional convolutional layer (Conv1d) has 64 filters, a kernel size of 16, stride 1, dilation 1, and uses a Rectified Linear Unit (ReLU) activation function. It is followed by a one-dimensional max-pooling layer (MaxPooling1d) with kernel size and stride 4. The second Conv1d layer has 128 filters, kernel size 16, stride 1, dilation 2, and ReLU activation function, followed by the same MaxPooling configuration as before. The third Conv1d layer has 256 filters, kernel size 16, stride 1, dilation 2, and ReLU activation function, again followed by the same MaxPooling1d. The fourth Conv1d layer has 512 filters, kernel size 32, stride 1, dilation 2, and ReLU activation function, followed by the same final MaxPooling1d layer.

The resulting CNN feature map is normalized using a layer normalization layer (LayerNorm) before being passed to the Transformer Encoder, providing pre-layer normalization. The feature map is reshaped such that the spacial dimension forms a sequence of 14 tokens, each represented by a 512-dimensional embedding. The Transformer stack consists of two encoder layers, each with an embedding dimension of 512 and four attention heads, a feed-forward hidden dimension of 2048, and a ReLU activation function in the feed-forward network. A dropout rate of 0.1 is applied within both the self-attention and feed-forward sublayers. Each TransformerEncoder layer also applies post-layer normalization by default.

Finally, the encoded representation is passed through three fully connected layers of 128, 64, and 6 neurons, respectively. The first two layers use ReLU activation functions, while the last layer outputs the six target parameters. A summary of the architecture and its parameter count is presented in~\autoref{table:proposed} and illustrated in ~\autoref{fig:hybrid-architecture}.

The convolutional layers are used to extract intricate local features from the time-series signal, benefiting from 1D-Convolutions that are specifically suited for temporal data. Through repeated convolution and pooling operations, a compressed yet informative feature representation is generated.

The Transformer layers are employed to capture long-range dependencies using self-attention, which assigns dynamic importance to different parts of the sequence. This mechanism enables the recognition of distinct phases of black hole coalescence, such as inspiral, merger, and ringdown, within the input signal. Unlike Recurrent Neural Networks, Transformers process the entire sequence in parallel, avoiding issues related to vanishing gradients or sequential bottlenecks.

\section{Results}
The results of this study are organized to demonstrate the performance and applicability of the proposed hybrid approach across both simulated and real gravitational-wave data. The section begins with a comparison against established CNN-based baselines on simulated datasets, followed by an evaluation on real events detected by LIGO/Virgo. Finally, posterior estimation using normalizing flows is presented, providing a first step toward uncertainty quantification and calibrated inference.

\subsection{Evaluation on simulated Data}
In this section, the proposed hybrid approach, which augments CNN-based models with Transformer Encoder layers is evaluated against three established CNN models: the ShallowModel and DeepModel ~\cite{george_deep_2018, george_deep_2018-1}, and the BNSModel ~\cite{krastev_detection_2021}. These models were originally developed for estimating the component masses of BBH and BNS systems, respectively, and were adapted in this study to predict six target parameters: the component masses and z-axis spin components of the two black holes, the luminosity distance to the gravitational wave event, and the inclination angle relative to the detector.

\begin{figure}[t!]
  \centering
  \includegraphics[width=0.9\linewidth]{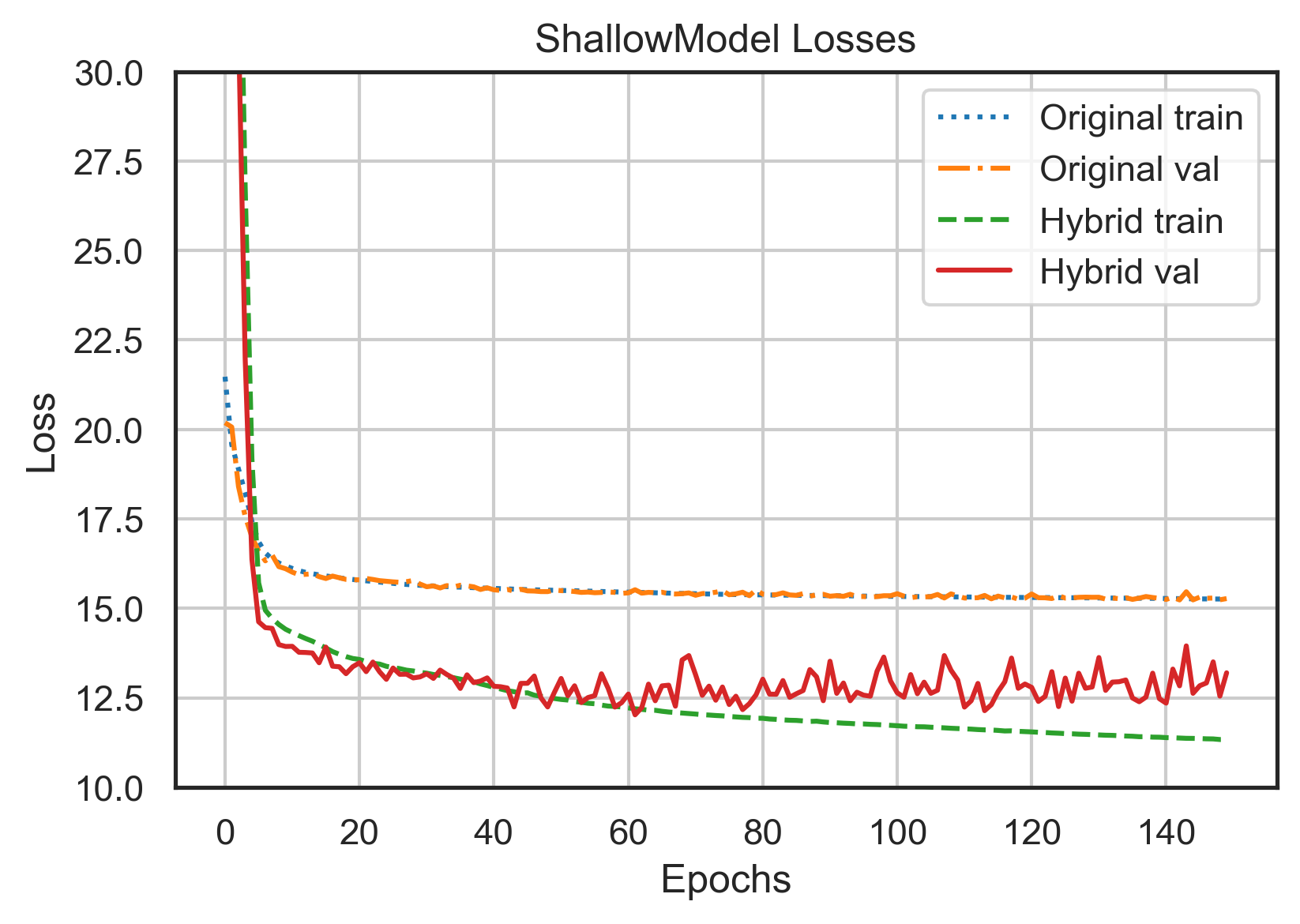}
  \caption{Training and Validation Loss per Epoch for the ShallowModel.}
  \label{fig:shallow_losses}
\end{figure}

\begin{figure}[t!]
  \centering
  \includegraphics[width=0.9\linewidth]{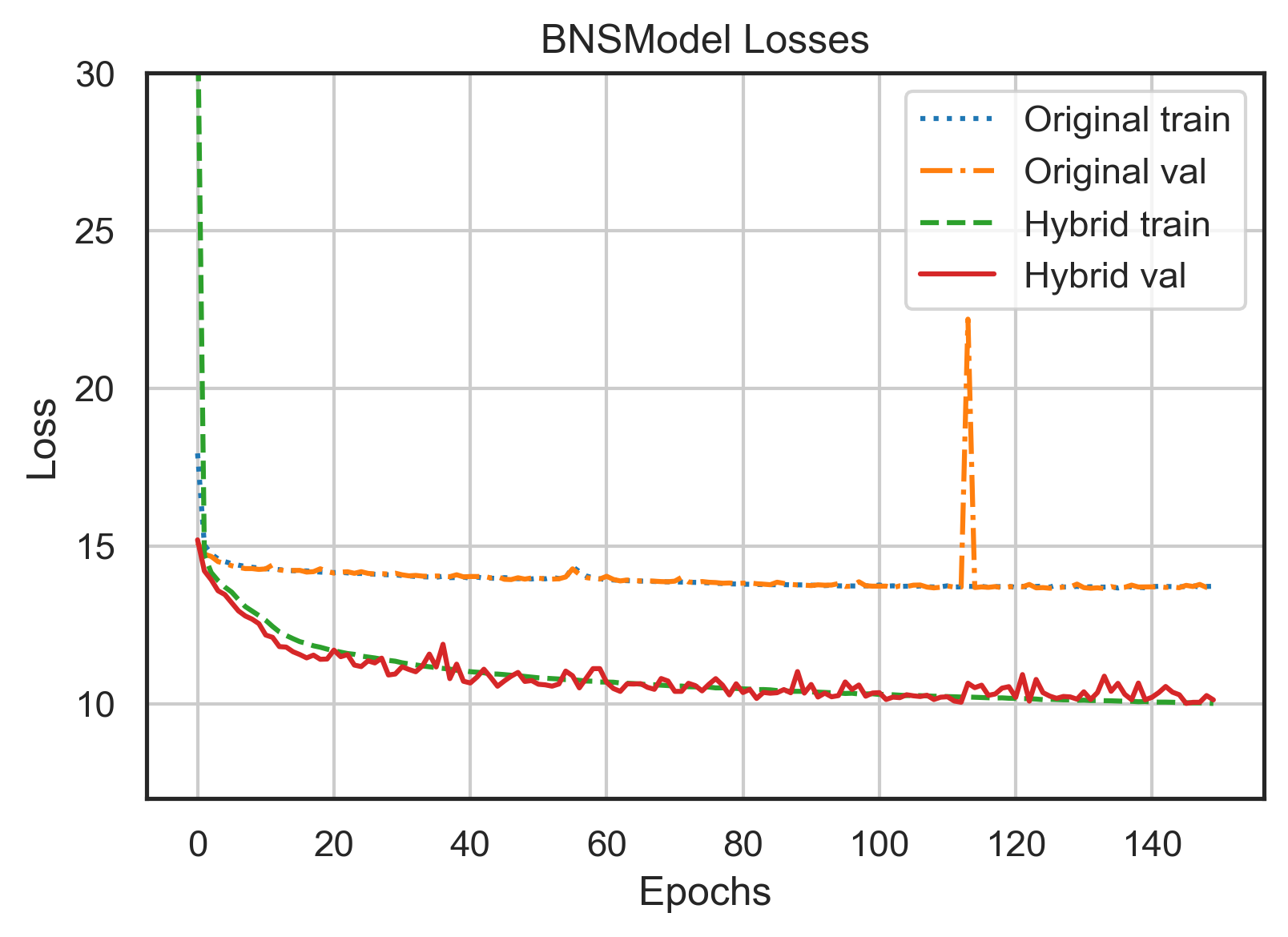}
  \caption{Training and Validation Loss per Epoch for the BNSModel.}
  \label{fig:bns_losses}
\end{figure}

\begin{figure}[t!]
  \centering
  \includegraphics[width=0.9\linewidth]{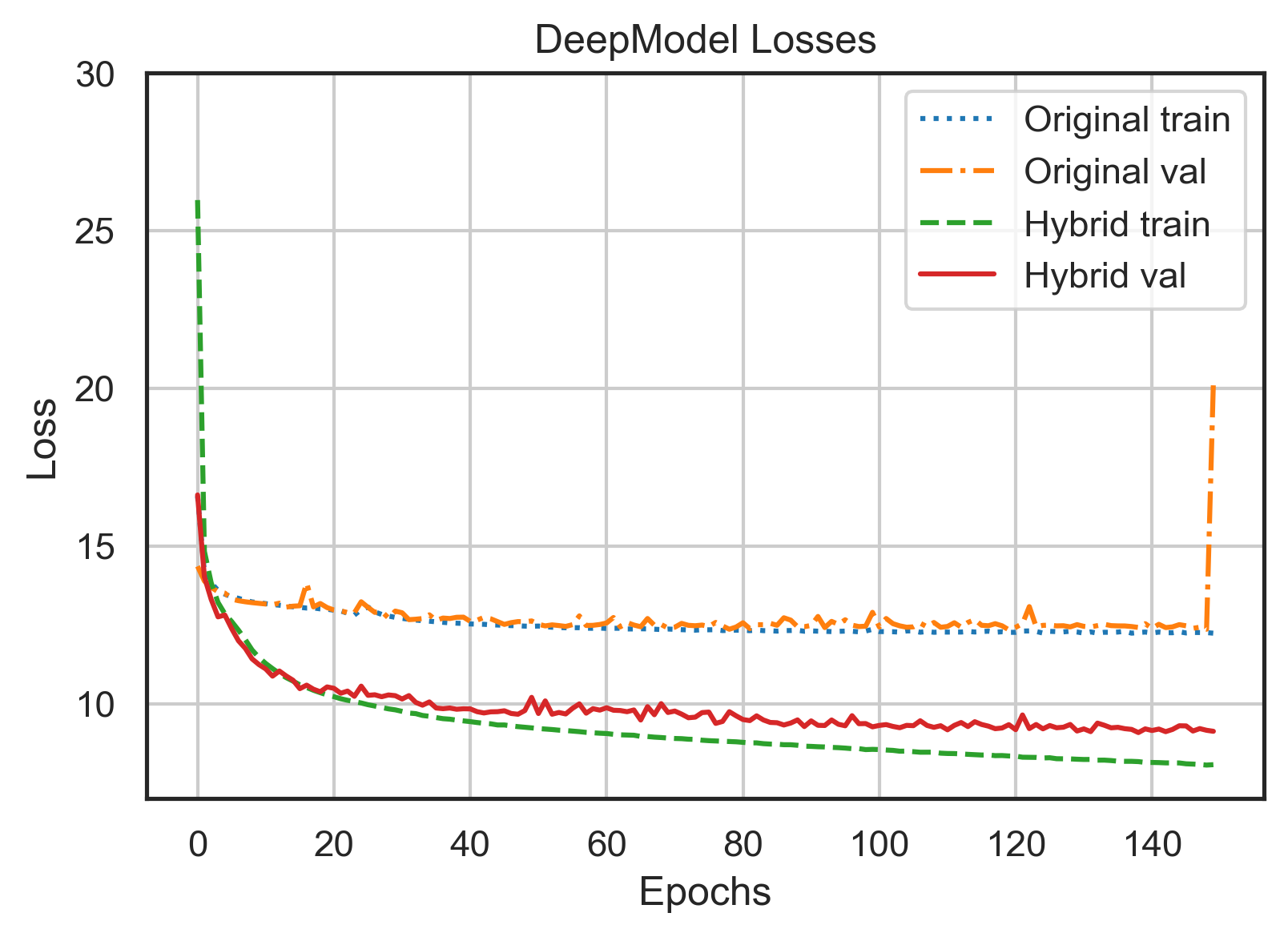}
  \caption{Training and Validation Loss per Epoch for the DeepModel.}
  \label{fig:deep_losses}
\end{figure}

All models were evaluated on gravitational wave signals embedded in Gaussian noise, simulating the noise characteristics of the LIGO detectors. This setting reflects the realistic conditions under which such parameter estimation must be performed in practice.

Training was conducted using time-domain gravitational wave strain data. The hybrid model was optimized using Adaptive Moment Estimation (Adam) optimizer with a learning rate of $10^{-4}$, while the baseline CNN models were trained with the same optimizer but a higher learning rate of $10^{-3}$, corresponding to their optimal performance settings. No learning rate scheduler was applied during training. All model were trained for 150 epochs with a batch size of 1,024 samples, utilizing an NVIDIA A100 80GB Graphics Processing Unit (GPU). This limit was set to manage the considerable computational cost, as training proved time-consuming across all architectures, even on high-performance hardware. Preliminary experiments indicated that model performance typically stabilized well before this point, making 150 epochs a practical compromise between efficiency and effectiveness. The Mean Absolute Error (MAE) was employed as the loss function for all models to ensure consistency in evaluation: 

\begin{equation}
\mathrm{MAE} = \frac{1}{n} \sum_{i=1}^{n} \left| y_i - \hat{y}_i \right|
\end{equation}

\noindent \Cref{fig:shallow_losses,fig:bns_losses,fig:deep_losses} illustrate the training and validation loss per epoch for the ShallowModel, BNSModel, and DeepModel, both in their original form and in their hybrid variant.
The evaluation metrics provided in ~\autoref{table:errors} were computed on the test data using the model weights from the epoch with the lowest validation loss, ensuring that the results reflect the model’s best generalization performance. Among these metrics, the Total Model Loss (TML) represents the mean of the Mean Absolute Error (MAE) across the six predicted parameters, and is defined as:

\begin{equation}
\mathrm{MAE}_{\text{Total}} = \frac{1}{n} \sum_{i=1}^{n} \left( \frac{1}{p} \sum_{j=1}^{p} \left| y_{i}^{(j)} - \hat{y}_{i}^{(j)} \right| \right)
\end{equation}

\begin{table*}[p]
    \caption{Error measurements for the ShallowModel, BNSModel, and DeepModel, with and without the addition of Transformer Encoders in the architecture, evaluated on Test data with Gaussian noise. A dash (–) indicates metrics that may be undefined due to parameters taking zero values.}

    \begingroup
    \setlength{\tabcolsep}{2pt}
    
    \begin{adjustbox}{width=1\textwidth}
        \centering
        \begin{tabular}{llcccccccc}
            \toprule                                                                                           
            {\textbf{Error}} & {\textbf{Parameter}} & \multicolumn{2}{c}{\textbf{ShallowModel}} &
            \multicolumn{2}{c}{\textbf{BNSModel}} &
            \multicolumn{2}{c}{\textbf{DeepModel}} & \\
            & & {Original} & {\textbf{Hybrid}} & {Original} & {\textbf{Hybrid}} & {Original} & {\textbf{Hybrid}} \\
            \midrule
            \multirow{6}{*}{\textbf{Mean Absolute Error (MAE)}}                
            & {Mass 1}             &{5.4772} & {\textbf{3.9700}} & {3.5765} & {\textbf{2.8303}} &{3.0745} & {\textbf{2.2820}} \\
            & {Mass 2}             &{4.9228} & {\textbf{2.5972}} & {3.2569} & {\textbf{1.9605}} &{2.5183} & {\textbf{1.8964}} \\
            & {Distance}           &{61.4424} & {\textbf{51.3420}} & {57.0403} & {\textbf{43.9518}} &{52.6632} & {\textbf{40.6117}} \\
            & {Inclination}        &{18.5598} & {\textbf{13.3030}} & {16.9829} & {\textbf{10.4960}} &{14.7085} & {\textbf{9.6311}} \\
            & {Spin 1}             &{0.3382} & {\textbf{0.2608}} & {0.3422} & {\textbf{0.1867}} &{0.3596} & {\textbf{0.2263}} \\
            & {Spin 2}             &{0.5895} & {\textbf{0.5170}} & {0.5868} & {\textbf{0.3941}} &{0.6034} & {\textbf{0.5273}} \\
            \hline
            \multirow{6}{*}{\textbf{Mean Absolute Percentage Error (MAPE)}}                
            & {Mass 1}             &{11.8023} & {\textbf{7.1373}} & {6.6356} & {\textbf{4.8146}} &{6.1071} & {\textbf{4.2950}} \\
            & {Mass 2}             &{20.7760} & {\textbf{8.5264}} & {12.0191} & {\textbf{6.2765}} &{9.1138} & {\textbf{6.2967}} \\
            & {Distance}           &{31.1254} & {\textbf{23.6762}} & {28.0905} & {\textbf{19.4288}} &{24.0225} & {\textbf{18.1377}} \\
            & {Inclination}        &{-} & {-} & {-} & {-} &{-} & {-} \\
            & {Spin 1}             &{-} & {-} & {-} & {-} &{-} & {-} \\
            & {Spin 2}             &{-} & {-} & {-} & {-} &{-} & {-} \\
            \hline
            \multirow{6}{*}{\textbf{Mean Absolute Relative Error (MARE)}}                
            & {Mass 1}             &{0.1180} & {\textbf{0.0714}} & {0.0664} & {\textbf{0.0481}} &{0.0611} & {\textbf{0.0429}} \\
            & {Mass 2}             &{0.2077} & {\textbf{0.0853}} & {0.1202} & {\textbf{0.0628}} &{0.0911} & {\textbf{0.0629}} \\
            & {Distance}           &{0.3112} & {\textbf{0.2368}} & {0.2809} & {\textbf{0.1943}} &{0.2402} & {\textbf{0.1813}} \\
            & {Inclination}        &{-} & {-} & {-} & {-} &{-} & {-} \\
            & {Spin 1}             &{-} & {-} & {-} & {-} &{-} & {-} \\
            & {Spin 2}             &{-} & {-} & {-} & {-} &{-} & {-} \\
            \hline
            \multirow{6}{*}{\textbf{Mean Relative Error (MRE)}}                
            & {Mass 1}             &{-0.0392} & {\textbf{-0.0236}} & {\textbf{-0.0071}} & {-0.0219} &{-0.0226} & {\textbf{0.0102}} \\
            & {Mass 2}             &{-0.0753} & {\textbf{0.0134}} & {-0.02616} & {\textbf{-0.0152}} &{-0.0128} & {\textbf{0.0092}} \\
            & {Distance}           &{-0.1631} & {\textbf{-0.1310}} & {-0.1490} & {\textbf{-0.1212}} &{-0.1131} & {\textbf{-0.1094}} \\
            & {Inclination}        &{-} & {-} & {-} & {-} &{-} & {-} \\
            & {Spin 1}             &{-} & {-} & {-} & {-} &{-} & {-} \\
            & {Spin 2}             &{-} & {-} & {-} & {-} &{-} & {-} \\
            \hline
            \multirow{6}{*}{\textbf{Mean Percentage Error (MPE)}}                
            & {Mass 1}             &{-3.9244} & {\textbf{-2.3628}} & {\textbf{-0.7072}} & {-2.1882} &{-2.2647} & {\textbf{1.0209}} \\
            & {Mass 2}             &{-7.5345} & {\textbf{1.3404}} & {-2.6157} & {\textbf{-1.5248}} &{-1.2773} & {\textbf{0.9294}} \\
            & {Distance}           &{-16.3141} & {\textbf{-13.1004}} & {-14.8966} & {\textbf{-12.1216}} &{-11.3150} & {\textbf{-10.9384}} \\
            & {Inclination}        &{-} & {-} & {-} & {-} &{-} & {-} \\
            & {Spin 1}             &{-} & {-} & {-} & {-} &{-} & {-} \\
            & {Spin 2}             &{-} & {-} & {-} & {-} &{-} & {-} \\
            \hline
            \multirow{6}{*}{\textbf{Root Mean Squared Error (RMSE)}}           
            & {Mass 1}             &{7.7020} & {\textbf{5.5651}} & {5.1051} & {\textbf{4.1662}} &{4.4048} & {\textbf{3.3689}} \\
            & {Mass 2}             &{6.6719} & {\textbf{3.7745}} & {4.3863} & {\textbf{3.1000}} &{3.4791} & {\textbf{2.7301}} \\
            & {Distance}           &{83.3083} & {\textbf{78.0487}} & {80.4025} & {\textbf{73.0403}} &{77.0995} & {\textbf{73.0750}} \\
            & {Inclination}        &{24.0310} & {\textbf{19.5472}} & {22.9772} & {\textbf{16.8205}} &{20.8715} & {\textbf{16.6720}} \\
            & {Spin 1}             &{0.4326} & {\textbf{0.3473}} & {0.4399} & {\textbf{0.2741}} &{0.4673} & {\textbf{0.3271}} \\
            & {Spin 2}             &{0.6986} & {\textbf{0.6187}} & {0.7005} & {\textbf{0.5051}} &{0.7198} & {\textbf{0.6367}} \\
            \hline
            \multirow{6}{*}{\textbf{Mean Squared Error (MSE)}}                
            & {Mass 1}             &{59.3210} & {\textbf{30.9703}} & {26.0623} & {\textbf{17.3570}} &{19.4018} & {\textbf{11.3497}} \\
            & {Mass 2}             &{44.5142} & {\textbf{14.2469}} & {19.2393} & {\textbf{9.6102}} &{12.1038} & {\textbf{7.4534}} \\
            & {Distance}           &{6940.2764} & {\textbf{6091.6044}} & {6464.5691} & {\textbf{5334.8863}} &{5944.3329} & {\textbf{5339.9509}} \\
            & {Inclination}        &{577.4873} & {\textbf{382.0946}} & {527.9511} & {\textbf{282.9299}} &{435.6178} & {\textbf{277.9568}} \\
            & {Spin 1}             &{0.1871} & {\textbf{0.1206}} & {0.1935} & {\textbf{0.0751}} &{0.2184} & {\textbf{0.1070}} \\
            & {Spin 2}             &{0.4880} & {\textbf{0.3828}} & {0.4907} & {\textbf{0.2552}} &{0.5182} & {\textbf{0.4054}} \\
            \hline
            \multirow{6}{*}{\textbf{R2 Score}}                                
            & {Mass 1}             &{0.8854} & {\textbf{0.9402}} & {0.9496} & {\textbf{0.9665}} &{0.9625} & {\textbf{0.9781}} \\
            & {Mass 2}             &{0.9139} & {\textbf{0.9725}} & {0.9628} & {\textbf{0.9814}} &{0.9766} & {\textbf{0.9855}} \\
            & {Distance}           &{0.6635} & {\textbf{0.7046}} & {0.6865} & {\textbf{0.7413}} &{0.7118} & {\textbf{0.7411}} \\
            & {Inclination}        &{0.3581} & {\textbf{0.5753}} & {0.4132} & {\textbf{0.6855}} &{0.5158} & {\textbf{0.6910}} \\
            & {Spin 1}             &{0.6254} & {\textbf{0.7585}} & {0.6126} & {\textbf{0.8496}} &{0.5628} & {\textbf{0.7859}} \\
            & {Spin 2}             &{0.0242} & {\textbf{0.2345}} & {0.0187} & {\textbf{0.4898}} &{-0.0360} & {\textbf{0.1894}} \\
            \midrule
            \multicolumn{2}{l}{\textbf{Total Model Loss (TML):}} & {15.2216} & {\textbf{11.9983}} & {13.6309} & {\textbf{9.9699}} & {12.3213} & {\textbf{9.1958}}  \\
            \bottomrule
        \end{tabular}
    \end{adjustbox}
    \endgroup
    \label{table:errors}
\end{table*}

A lower loss value indicates better model performance for this task. As shown in \Cref{fig:shallow_losses,fig:bns_losses,fig:deep_losses} and Table~\ref{table:errors}, the hybrid variants of the three baseline CNN models consistently achieve lower Total Model Loss (TML) compared to their original forms when evaluated on gravitational wave signals embedded in Gaussian noise. These results demonstrate that integrating Transformer encoder layers improves predictive accuracy under realistic noise conditions across all evaluated architectures.

In terms of computational efficiency, the DeepModel and its hybrid variant, despite being the most parameter-rich and potentially the slowest architectures, achieve average inference times of approximately 36ms and 40ms per strain, respectively. Notably, the hybrid model incurs only a marginal increase in latency compared to its baseline, while both remain orders of magnitude faster than traditional Bayesian inference methods.

Mean Absolute Error (MAE), Root Mean Square Error (RMSE), and Mean Squared Error (MSE) quantify prediction accuracy, with RMSE and MSE penalizing larger errors more heavily, and MSE being especially sensitive to outliers. Mean Absolute Percentage Error (MAPE) and Mean Percentage Error (MPE) express errors as percentages, making them useful for parameters with varying scales. Mean Absolute Relative Error (MARE) and Mean Relative Error (MRE) normalize errors by true values, providing scale-invariant evaluation—particularly relevant for mass, and distance. MPE and MRE can also indicate systematic bias, where consistently positive or negative values reveal over- or underestimation, while values near zero suggest the model maintains minimal bias. The R² score measures the proportion of variance explained by the model, with values near 1 indicating a good fit, and negative values suggesting worse performance than simply predicting the mean.

Some of the evaluation metrics used in this study, specifically MAPE, MPE, MRE, and MARE, become undefined when the true target value is zero, due to division by zero. In such cases, the corresponding entries in the results tables are denoted with a dash (–), as they do not provide valid statistical information. This situation primarily occurs for the inclination angle and the component spins along the z-axis. In contrast, metrics such as MAE, MSE, RMSE, and R², which do not involve division by the target value, remain unaffected by zeros and provide complementary insights into model performance.

As shown in Table~\ref{table:errors}, lower MAE and RMSE values were achieved by the proposed approach, indicating improved predictive accuracy. Reductions in MAPE and MARE further reflect enhanced relative performance across target parameters with varying numerical scales. Although the MPE and MRE values are not uniformly close to zero, they remain consistently lower than those of the benchmark models, suggesting a reduced bias toward systematic over- or underestimation. This is particularly significant for astrophysical quantities such as luminosity distance, where biased predictions may lead to misleading physical interpretations. Additionally, a higher R² score indicates that a greater proportion of the variance in the data was captured by the proposed approach compared to the CNN-based baselines. These findings indicate that the integration of Transformer Encoder layers enhances the model's capacity to extract meaningful patterns from the data, resulting in superior generalization and overall performance.

Notice that during training, one of the three original CNN models exhibited a brief spike in validation loss at a single epoch, while another model showed a transient increase during the final epoch. These deviations are likely due to stochastic fluctuations in weight updates or temporary overfitting to specific mini-batches. Such behavior is not uncommon in Deep Learning models, especially when trained with adaptive optimizers, and does not necessarily imply poor generalization. The stability of validation loss before and after these fluctuations indicates that the models quickly recovered and maintained consistent performance.

The three baseline CNN models consist of one-dimensional convolutional layers followed by fully connected layers. While these architectures have shown strong performance in estimating intrinsic parameters, such as component masses, their limitations become more apparent when extended to the joint estimation of six parameters, including spin components, luminosity distance, and inclination angle. In the proposed hybrid approach, the baseline CNN architectures remain unchanged, with the sole modification being the insertion of Transformer Encoder layers after the convolutional feature extractor. This design preserves the local feature extraction capability of CNNs while enhancing them with attention-based global context modeling. The convolutional layers are responsible for extracting compact and denoised local waveform features, while the Transformer layers dynamically prioritize the most informative time steps through self-attention.

This hybrid approach enables the models to better differentiate between waveform patterns generated by varying physical parameters, thus improving estimation precision. Collectively, the findings support the hypothesis that integrating convolutional and Transformer-based mechanisms leads to more accurate and noise-resilient predictions of black hole binary parameters from gravitational wave signals.

\begin{figure}[t!]
  \centering
  \includegraphics[width=1\linewidth]{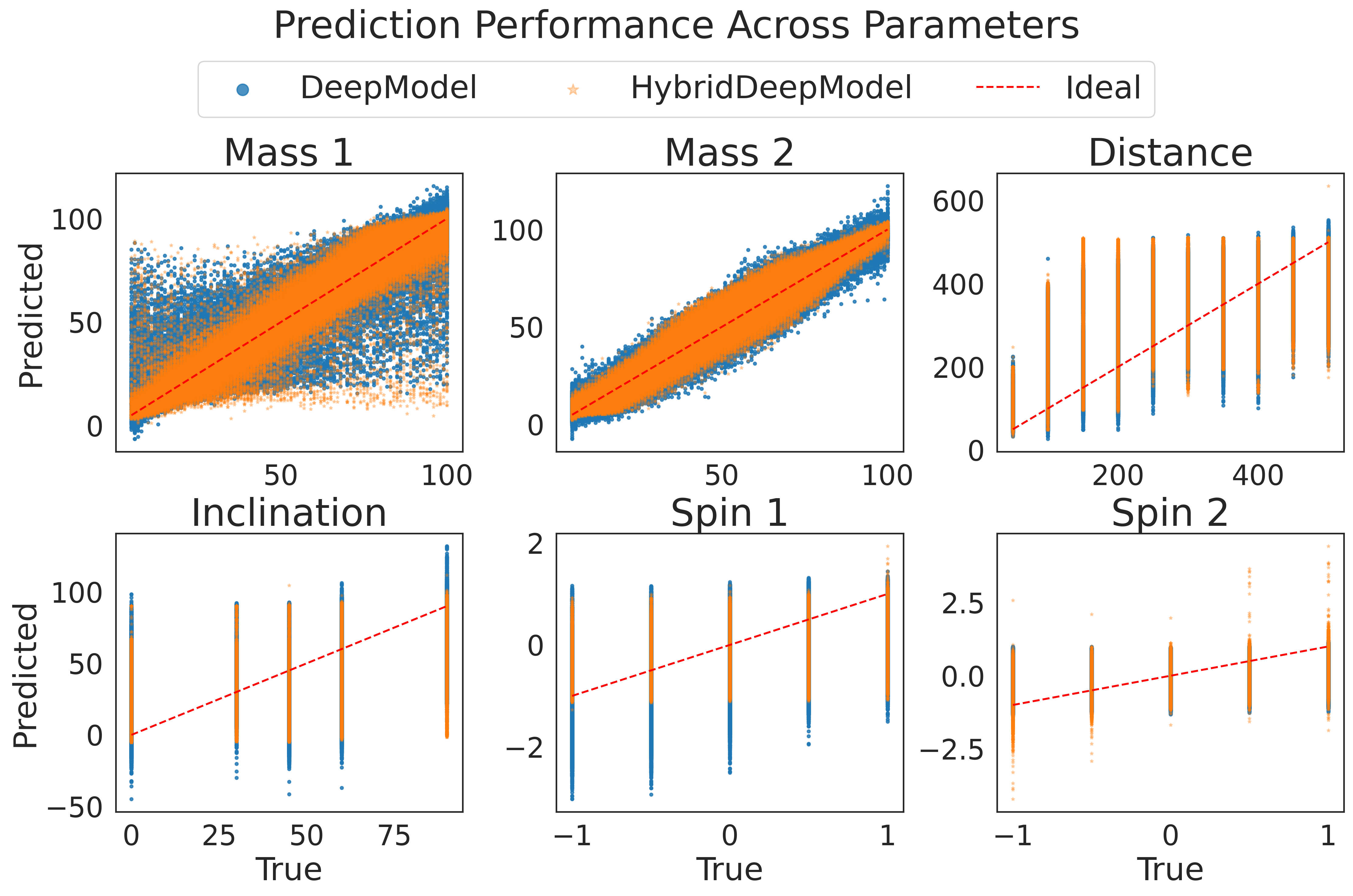}
  \caption{Comparison between true and predicted physical parameters for simulated gravitational wave signals. Each subplot shows the predicted versus true values for the component masses ($m_1$, $m_2$), luminosity distance ($d_L$), inclination ($\iota$), and spin components along the z-axis ($\chi_{z_{1}}$, $\chi_{z_{2}}$). Results are shown for the DeepModel (blue) and DeepModel-hybrid (orange). The red dashed line represents the ideal one-to-one correlation.}
  \label{fig:true_vs_predicted}
\end{figure}

~\autoref{fig:true_vs_predicted} presents a direct comparison between the true and predicted values of the six physical parameters estimated from the simulated gravitational wave signals. The DeepModel-hybrid exhibits markedly improved performance over the baseline DeepModel, as indicated by its tighter clustering around the ideal one-to-one line across all parameters. The improvement is particularly evident in the estimation of the component masses ($m_1$, $m_2$) where the hybrid architecture achieves more stable and accurate predictions. The luminosity distance ($d_L$), inclination ($\iota$), and spin parameters ($\chi_{z_{1}}$, $\chi_{z_{2}}$), show larger intrinsic variance due to their weaker imprint on the strain amplitude, however, the hybrid model still produces a visibly more consistent trend. These results highlight the advantage of integrating Transformer encoder layers into the CNN baseline, enabling the model to capture both local and global dependencies in the waveform morphology more effectively.

To reinforce the claim that the addition of Transformer Encoder layers is the key factor behind the improved point estimation of the hybrid approach, t-SNE visualizations of the latent representations were generated for the six target parameters. Due to space limitations, only results for the combinations of $m_1$ and $m_2$ are shown, though similar trends were observed for all the other parameters.

\begin{figure}[t!]
  \centering
  \includegraphics[width=1\linewidth]{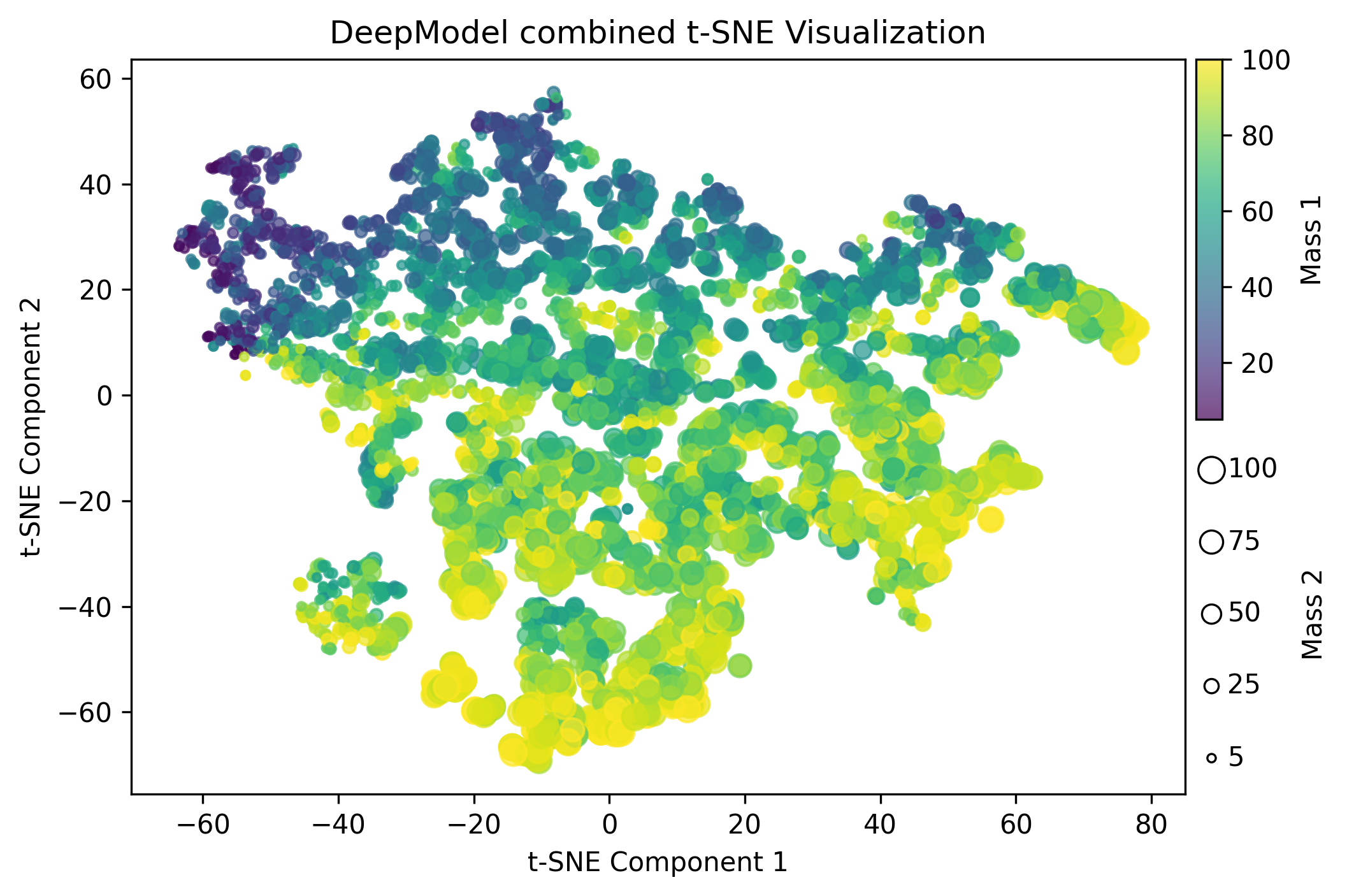}
  \caption{t-SNE visualization of latent representations from the first fully connected layer of DeepModel. Point color encodes $m_1$ and size encodes $m_2$. Overlapping regions indicate weaker separation of mass-related features.}
  \label{fig:t-sne_deep}
\end{figure}

\begin{figure}[t!]
  \centering
  \includegraphics[width=1\linewidth]{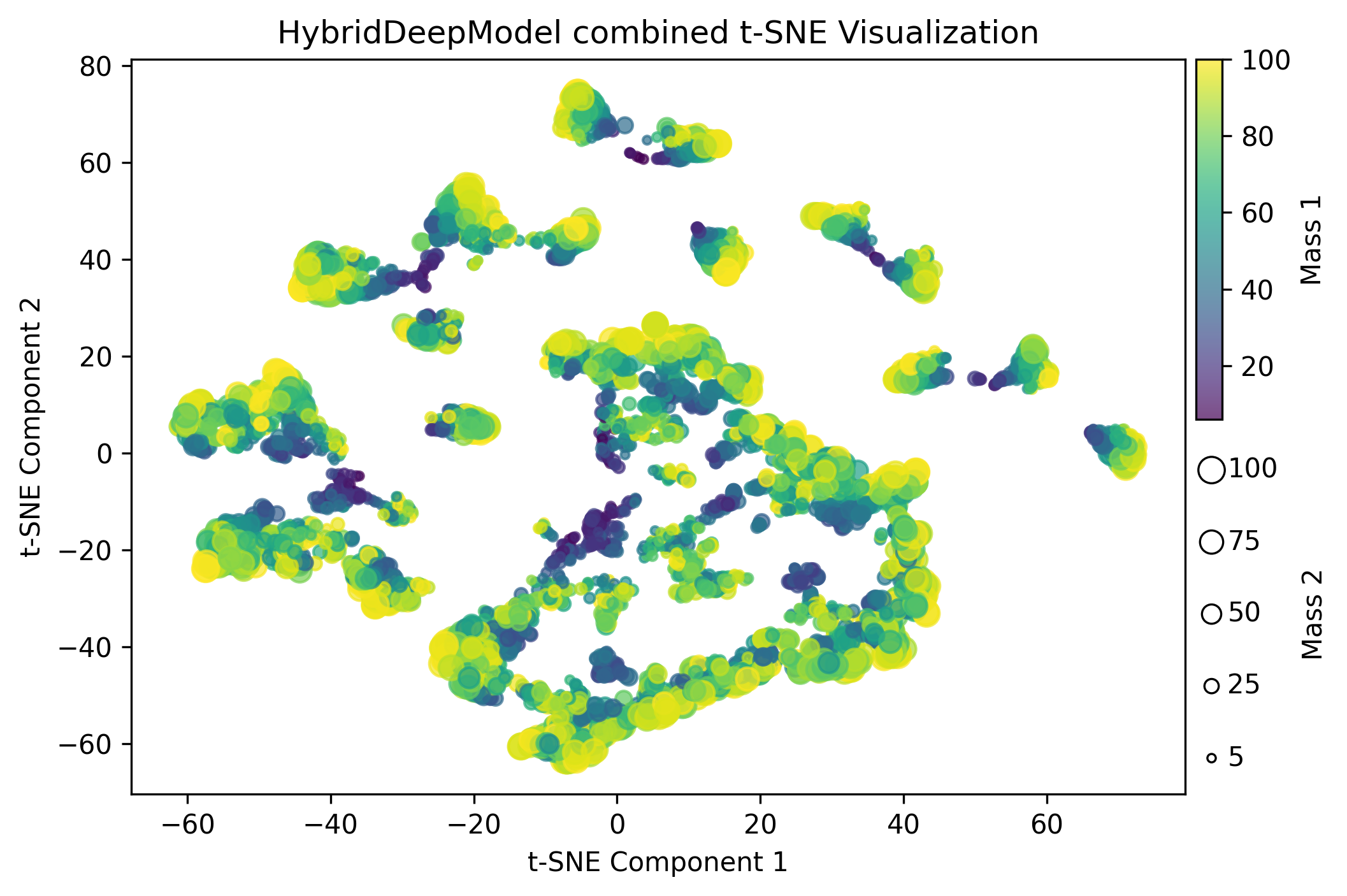}
  \caption{t-SNE visualization of latent representations from the first fully connected layer of DeepModel-hybrid. Point color encodes $m_1$ and size encodes $m_2$. Distinct clusters and smooth transitions indicate a more structured latent representation.}
  \label{fig:t-sne_deep_hybrid}
\end{figure}

The \Cref{fig:t-sne_deep,fig:t-sne_deep_hybrid} illustrate how DeepModel and DeepModel-hybrid encode mass-related features in a reduced two-dimensional space, providing insight into each model's ability to separate different parameter values. The hybrid model yields a clearer separation of data clusters in the latent space compared to the CNN-only baseline, indicating that the Transformer layers enhance the discriminative structure of the learned feature space. This improved organization in the latent representation likely underlies the superior quantitative performance of the hybrid model, as it suggests a more effective encoding of parameter-dependent variations within the gravitational wave signals. This qualitative evidence therefore supports the quantitative improvements observed in the evaluation metrics, further confirming that the Transformer Encoder contributes critically to the success of the hybrid architecture.

\subsection{Evaluation on Real Gravitational-Wave Events}
To further assess the generalization ability of the proposed approach, both the best performing CNN baseline (DeepModel) and its hybrid extension were evaluated on real gravitational-wave strain data from the GWOSC catalog. For consistency with the training domain, the test set was restricted to BBH events whose component masses and luminosity distances lie within the parameter ranges used in the artificially generated training data. Since only five events (GW170608, GW151226, GW150914, GW200202\_154313, GW191216\_21338) satisfied these conditions, two fine-tuning strategies were examined. In the first strategy, both models were first fine-tuned for 100 epochs on a single event and subsequently tested on the remaining four, with the procedure repeated five times so that each event was used once for fine-tuning. In the second one, four events were used for fine-tuning and the remaining event was reserved for testing. Corresponding values for the inclination angle of the BBH system and the individual black hole spins are not provided in the public event catalog and therefore could not be included in this validation.

The performance of the two models on real GWOSC data was quantified using MAE. To account for catalog uncertainties, a bound-based criterion was adopted, in which predictions that fell within the reported lower and upper bounds of the catalog parameters were treated as exact and assigned zero error. Under this criterion, for the single-event fine-tuning strategy (first strategy), where each model was fine-tuned for 100 epochs on one event and tested on the remaining four, the mean MAE across the five runs was 9.1688 $\pm$ 6.3576 for the hybrid model, and 415.3885 $\pm$ 433.3958 for the baseline model. For the four-event fine-tuning strategy (second strategy), where four events were used for fine-tuning and the remaining event for testing, the mean MAE across the five runs was 7.4690 $\pm$ 4.1364 for the hybrid model and 199.6947 $\pm$ 145.2793 for the baseline model. 

When evaluated without applying this bound-based correction, the mean MAE increased to 28.5177 $\pm$ 5.8427 and 416.1169 $\pm$ 432.8633 for the single-event fine-tuning, and 11.3361 $\pm$ 2.6519 and 206.9566 $\pm$ 161.2731 for the four-event fine-tuning, for the hybrid and baseline models respectively. These results indicate that incorporating additional events during fine-tuning leads to more robust generalization on unseen data. In both strategies, the hybrid model consistently outperformed the baseline CNN, demonstrating improved parameter estimation accuracy. Notably, although the models were originally trained only on artificially generated signals with Gaussian noise, the hybrid architecture shows promising generalization potential to real GWOSC data, whereas the baseline CNN showed substantially poorer performance under the same conditions. Detailed evaluation results for both fine-tuning strategies can be found in Tables A1 and A2 (Supplementary Material).

It should be noted that with only five real events, the reported mean MAE and standard deviations should be interpreted with caution, as these statistics are not robust. While the results are promising, a larger sample of real events would be necessary to draw definitive conclusions about model performance in realistic observational scenarios.

\subsection{Posterior Estimation with Normalizing Flows}
To further demonstrate the superior feature representations learned by the proposed hybrid approach and extend the capability of the models beyond point estimation, a normalizing flow (NFlow) was incorporated on top of the hybrid and baseline CNN architectures. This addition allows the models to estimate full posterior distributions of the target parameters, providing a measure of uncertainty in the predictions. By learning a bijective mapping from a simple base distribution to the complex posterior, the NFlow-based models are able to capture correlations and multi-modal structures in the parameter space that are inaccessible to deterministic point-estimation networks. This extension thus enables a more complete probabilistic inference framework for gravitational-wave parameter estimation, while building upon the previously demonstrated improvements of the hybrid architecture.

To evaluate the posteriors obtained with this extension, the original model weights were frozen, and only the NFlow component was trained using the negative log-likelihood (NLL) loss. This frozen-weight strategy ensures a controlled comparison between the best CNN-only baseline (DeepModel) and the Hybrid extension, so that differences in posterior quality arise from the learned features rather than from end-to-end co-adaptation between model weights and the flow. The quality of the resulting posteriors was assessed using per-parameter and combined p-values, as well as mean Continuous Ranked Probability Score (CRPS) and sharpness. The mean CRPS reflects both the accuracy and calibration of the posterior, with lower values indicating better agreement with the true parameters. Sharpness quantifies the concentration of the posterior distributions, where lower values correspond to narrower and more confident predictions. Finally, the combined p-value evaluates overall calibration across all parameters, with values close to 0.5 indicating well-calibrated posteriors and values near 0 or 1 signaling under- or over-dispersion. As shown in Table~\ref{table:posteriors}, the hybrid model achieves substantially more reliable calibration than the baseline CNN, with a combined p-value closer to the ideal value of 0.5, while also yielding considerably lower CRPS and sharper posteriors. These results indicate that the hybrid approach provides both more precise and better-calibrated posterior estimates than the CNN-only baseline. 

\begin{figure}[h!]
  \centering
  \includegraphics[width=0.9\linewidth]{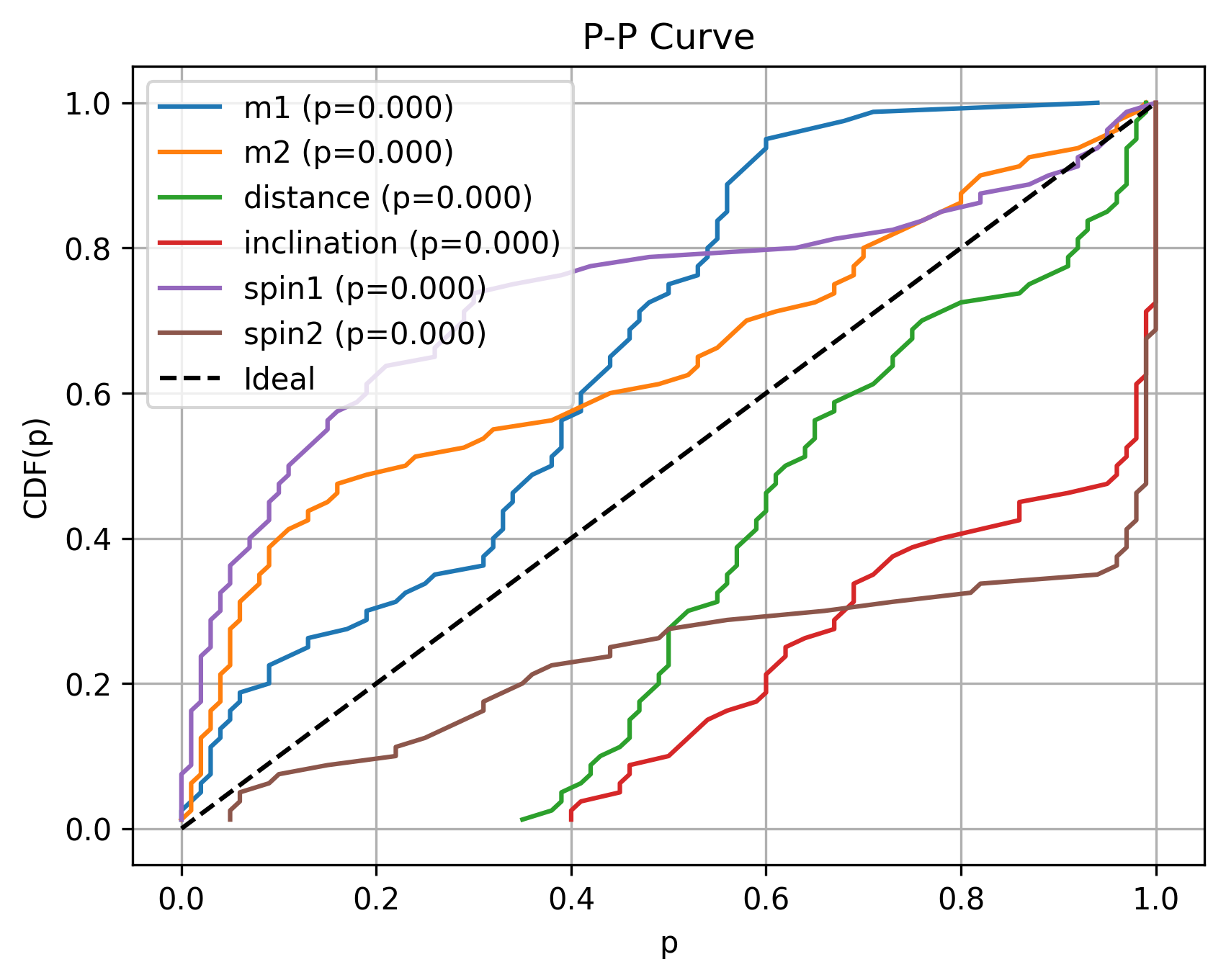}
  \caption{P-P plot for the DeepModel-Flow. The legend shows the p-values of the individual parameters, with a combined p-value of 1.2858e-75.}
  \label{fig:deepflow_ppcurve}
\end{figure}

\begin{figure}[h!]
  \centering
  \includegraphics[width=0.9\linewidth]{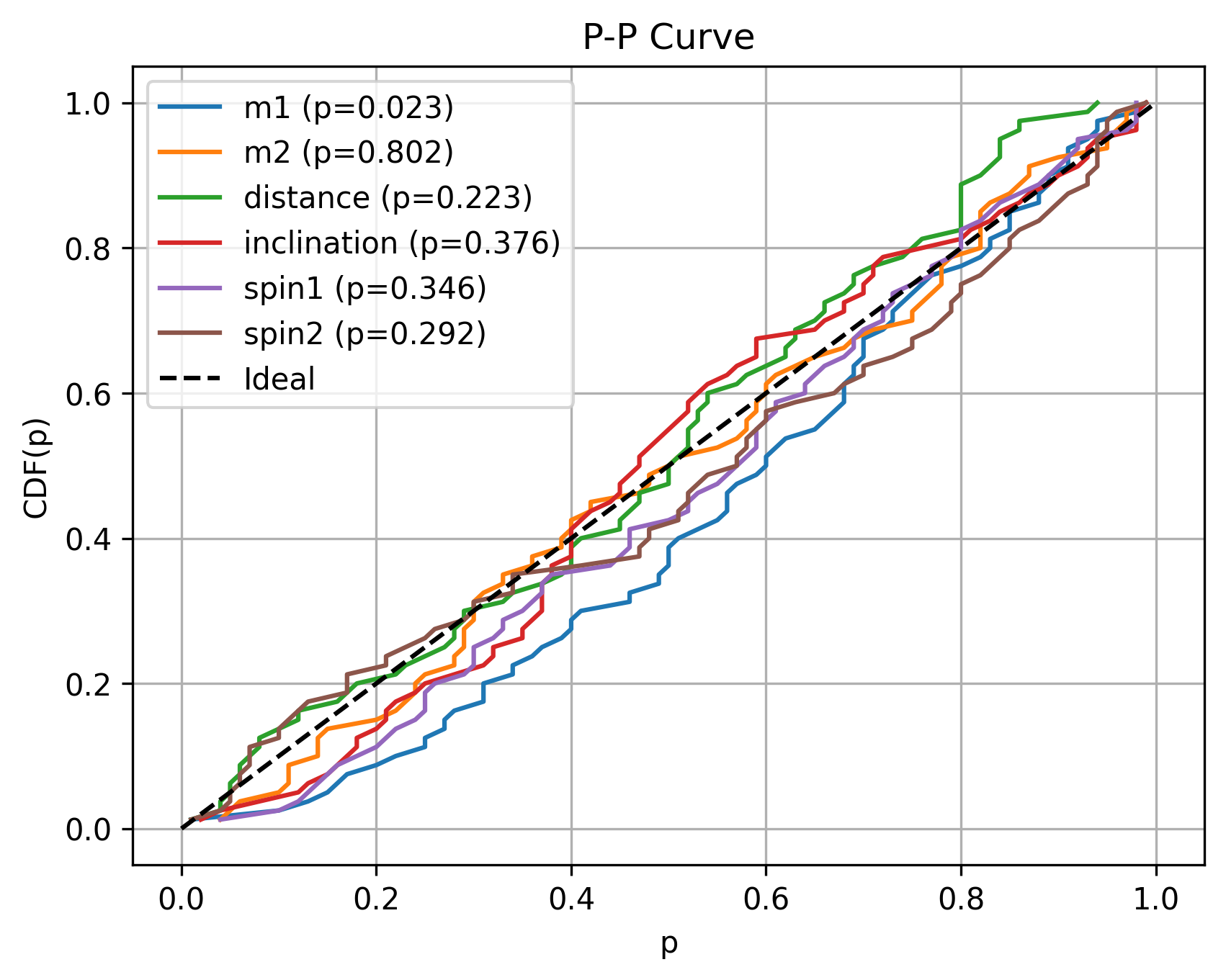}
  \caption{P-P plot for the HybridDeepModel-Flow. The legend shows the p-values of the individual parameters, with a combined p-value of 0.1296.}
  \label{fig:hybridflow_ppcurve_frozen}
\end{figure}

\begin{table}[h!]
\centering
\caption{Posterior quality metrics for the baseline CNN and the hybrid CNN-Transformer models.}
\resizebox{\linewidth}{!}{%
    \begin{tabular}{lrrr}
        \multicolumn{4}{c}{\textbf{Posterior evaluation metrics}}                      \\ \hline
        {\textbf{Model}}                & {\textbf{Mean CRPS}}  & {\textbf{Mean Sharpness}}  & {\textbf{Combined P-value}} \\ \hline
        {\textbf{DeepModel-Flow}}& {3.2982e+14}                 & {2.3052e+15}                     & {1.2858e-75}         \\
        {\textbf{HybridDeepModel-Flow}} & {6.6938}      & {13.0738}                 & {0.1296}             \\ \hline
        
    \end{tabular}
}
\label{table:posteriors}
\end{table}

\section{Summary and conclusions}
By combining local feature extraction via convolutional layers with global sequence modeling through Transformer encoders, a hybrid approach has been presented for the estimation of gravitational-wave parameters. This approach enables the capture of complex, multi-scale patterns within gravitational-wave signals, which is essential for distinguishing the subtle signatures of both intrinsic and extrinsic source characteristics.

Empirical results have shown that the proposed hybrid approach, in which the original CNN-based benchmark models are modified solely by the addition of Transformer Encoder layers, consistently outperforms their original unmodified counterparts across a range of evaluation metrics in noise-injected scenarios. Lower prediction errors, reduced estimation bias, and higher R² scores have been observed, reflecting improved generalization and robustness. Notably, enhanced performance has been achieved for particularly challenging parameters such as luminosity distance and inclination angle. 

This study offers several notable contributions to the field of gravitational-wave parameter estimation. First, it presents a systematic evaluation demonstrating that the integration of Transformer encoder layers into established CNN architectures yields consistent improvements across multiple performance metrics. Second, the proposed approach is validated not only on simulated data but also on a limited set of real gravitational-wave events, underscoring its practical relevance. Third, the potential of the proposed approach is further examined by integrating NFlows atop a frozen feature extractor. This design enables posterior inference and illustrates that the learned features support principled uncertainty quantification. Fourth, the approach exhibits a substantial computational efficiency advantage over traditional Bayesian inference techniques such as MCMC and nested sampling, aligning with the broader motivation to develop scalable deep learning-based alternatives for gravitational-wave data analysis.

As gravitational-wave detectors advance in sensitivity and the volume of detected events grows, reliable and accurate parameter estimation strategies such as the one proposed here will become increasingly important. This hybrid approach is expected to play a key role not only in characterizing individual events but also in enabling population-level studies and stringent tests of general relativity in the strong-field regime. 

Nonetheless, several key limitations remain. A primary limitation of this work is the exclusive use of simulated Gaussian noise, which does not capture the non-Gaussian, non-stationary glitches present in real detector data. This restricts the realism of the training environment and may limit generalization to actual observations. Additionally, the current model estimates only six parameters and does not incorporate effects such as spin precession, higher-order modes, or sky localization, thereby limiting its applicability to a subset of BBH events. Validation on real gravitational-wave events was performed on a very small sample, and broader testing is required to confirm robustness and generalizability. Furthermore, while the point estimation approach demonstrates strong performance, it does not natively provide uncertainty quantification. This drawback was partially addressed through the integration of NFlows, but further developments in simulation-based inference, along with validation on real gravitational-wave events, as demonstrated in previous works \cite{Green_2020, green2020completeparameterinferencegw150914, Dax_2021, Dax_2025},  are necessary to provide more reliable uncertainty quantification and credible intervals.

Future work should aim to address these limitations by incorporating real detector noise and glitch artifacts, expanding the parameter space to include additional astrophysical quantities, and performing systematic validation on a larger set of real events. Finally, while systematic ablation studies are required to fully isolate the specific contribution of the Transformer encoder relative to CNN-only or flow-based baselines, the results presented in this work consistently indicate that the inclusion of Transformer encoder layers enhances the performance of CNN architectures. This suggests that the observed improvements are closely linked to the attention-based context modeling introduced by the Transformer component. Extensions along these directions, together with adaptation to multi-detector or multi-messenger frameworks, could ultimately contribute to a more comprehensive understanding of compact binary coalescences.

\bibliographystyle{elsarticle-num-custom}

\bibliography{references}

\end{document}